\documentstyle[twocolumn,psfig,twoside,twocolumn]{article}
\begin{document}
\setlength{\textwidth}{18cm}
\newcommand{\ie}{{\em i.e. }}
\newcommand{\al}{{\em et al.}}
\newcommand{\eg}{{\em e.g. }}
\newcommand{\X}{\mbox{\boldmath X}}
\newcommand{\beq}{\begin{equation}}
\newcommand{\eeq}{\end{equation}}
\newcommand{\Taver}[1]{  \left \langle #1 \right \rangle }

\title{Coupled Two-Way Clustering Analysis of Breast Cancer and Colon Cancer
Gene Expression Data}
\author{Gad Getz$^1$, Hilah Gal$^1$, Itai Kela$^1$, Dan A. Notterman$^2$
and Eytan Domany$^1$ \\
$^1$Department of Physics of Complex Systems, Weizmann Institute
of Science, Rehovot 76100, Israel\\
$^2$Department of Molecular Biology, Princeton University,
Princeton, NJ 08544 and \\ Department of Pediatrics, Robert Wood
Johnson Medical School, New Brunswick, NJ }
\date{\today}
\maketitle

\noindent
{\bf ABSTRACT}\\
We present and review Coupled Two Way Clustering, a method
designed to mine gene expression data. The method identifies
submatrices of the total expression matrix, whose clustering
analysis reveals partitions of samples (and genes) into
biologically relevant classes. We demonstrate, on data from colon
and breast cancer, that we are able to identify partitions that
elude standard clustering analysis.\\
{\bf Availability:} Free, at
http://ctwc.weizmann.ac.il.\\
{\bf Contact:} eytan.domany@weizmann.ac.il \\
{\bf Supplementary Information:}
http://www.weizmann.\\ac.il/physics/complex/compphys/bioinfo2/ \\

\noindent
\section*{INTRODUCTION}
Two nearly concurrent recent advances - the development of high density DNA chips
and the deciphering of the human genome - hold great promise for significant progress in
biomedical research. A large umber of studies have been published within the last years,
attempting to classify, explain and perhaps help cure several human diseases,
on the basis of gene expression
levels measured for populations of diseased and healthy subjects. Different
forms of cancer have been at the focus of such studies from early on, using all
available
chip technologies.

A DNA chip measures simultaneously the expression levels of
thousands of genes for a particular sample. Since a typical
experiment on human subjects provides the expression profiles of
several tens of samples (say $N_s \approx 100$), over several
thousand ($N_g$) genes whose expression levels passed some
threshold, the outcome of such an experiment contains between
$10^5$ and $10^6$ numbers. These are summarized in an $N_g \times
N_s$ {\it expression table}; each row corresponds to one
particular gene and each column to a sample, with the entry
$E_{gs}$ representing the expression level of gene $g$ in sample
$s$. Analysis of such massive amounts of data poses a serious
challenge for the development and application of novel
methodologies.

We present here {\it Coupled Two Way Clustering} (CTWC), a
recently introduced  method (Getz \al, 2000), designed to "mine"
gene expression data, and demonstrate its strength by applying it
to breast cancer and colon cancer data. The CTWC software is
accessible at http://ctwc.weizmann.ac.il (Getz and Domany 2002).

CTWC is based on {\it clustering}, and as such it is {\it
unsupervised} and capable of discovering unanticipated partitions
of the data, exploring its structure on the basis of correlations
and similarities that are present in it. In the context of gene
expression, such analysis has two obvious goals:\\
1. Find groups of genes that have correlated expression profiles.
The members of such a group may take part in the same biological
process.\\
2. Divide the tissues into groups with similar gene expression
profiles. Tissues that belong to one group are expected to be in
the same biological (e.g. clinical) state.\\
The straightforward way to carry out such analysis is to cluster
the data in {\it two ways}. Denote the set of all genes that
passed a threshold by $ G1$ and the set of all samples by $S1$.
Each gene is a point in an $\vert S1 \vert$ dimensional space; the
first clustering operation, $G1(S1)$, clusters all genes on the
basis of their expression levels over all samples. The
complementary operation, $S1(G1)$, clusters the samples on the
basis of their expression levels over all $\vert G1 \vert$ genes.
A variety of clustering methods have been used to perform these
operations. Clustering is based on some measure of similarity of
pairs of samples $s,s'$ which, in turn, is governed by their
"distance" in the $\vert G1 \vert$ dimensional space of expression
levels.

As several groups noticed (Perou \al, 2000; Cheng and Church,
2000; Califano \al, 2000), one runs into a severe difficulty with
this simple "all against all" clustering approach. The reason is
that in general only a small subset of $N_r$ relevant genes is
involved in one particular biological process of interest. Since
usually $N_r << |G1|$, the "signal" provided by this subset may be
completely masked by the "noise" generated by the much larger
number of the other genes. Furthermore, it may well happen that in
order to assign samples into two clinically meaningful classes
(e.g. adenoma and carcinoma) on the basis of the $N_r$ relevant
genes, one must first remove a previously identified group of
samples (e.g. healthy tissue), and cluster only the remaining
$N_s'<N_s$ tumors (using only the $N_r$ relevant genes). Thus one
should look for special $N_r \times N_s'$ {\it submatrices} of the
total expression matrix; such a search is problematic since an
exhaustive enumeration of such submatrices is of exponential
complexity. CTWC provides a heuristic method to search for such
submatrices. It has been used successfully to mine data (Getz \al,
2000) from experiments on colon cancer (Alon \al, 1999) and
leukemia (Golub \al, 1999), glioblastoma (Godard \al, 2002),
breast cancer (Kela 2001) and antigen chips (Quintana \al, 2002).
We present here results obtained by a new, more interactive usage
of CTWC on cDNA microarray data from breast cancer (Perou \al
2000, referred to as {\bf PAL}; Sorlie \al, 2001, referred to as
{\bf SAL}) and on oligonucleotide microarray data from colon
cancer patients (Notterman \al, 2001).

The analysis of Notterman \al  stopped at two way clustering,
which is the first step of CTWC - here our aim is to demonstrate
that by going beyond this step we uncover new partitions of the
samples. The situation with the breast cancer data is more
interesting. PAL noticed that simple two way clustering did not
partition the samples in a meaningful way, and pruned their
original set of $\vert G1 \vert = 1753$ down to 496 "intrinsic
genes", that were selected in a knowledge based way (which can be
applied only if the data contains pairs of samples taken from the
same patients). CTWC also identifies (much smaller) sets of genes
that are used to cluster the samples, but it is done in an
automated, objective, generally applicable  way. It was not clear
a priori that CTWC will reproduce the valuable observations of PAL
and SAL, and even less that it will yield new results of possible
biological or clinical significance.

\section*{MATERIALS AND METHODS}

{\bf Expression data - breast cancer.} We studied two data sets on
breast cancer. The first expression matrix was measured and
analyzed by PAL and the second by SAL. The PAL study characterizes
gene expression profiles of 84 samples (the set $S$), composed of
65 tumors (sample set $S1$) and 19 cell lines, using cDNA
microarrays, representing 8,102 human genes. Twenty of the 65
tumors were sampled twice; 18 from patients who were treated with
doxorubicin (chemotherapy) for an average of 16 weeks, with
surgical biopsy done {\it before} and {\it after} the treatment,
and two more tumors were paired with a lymph node metastasis from
the same patient. The 25 remaining specimens included 22 tumors
and three samples from normal breast tissues (nevertheless, we
refer to these also as "tumors"). The full expression matrix
included 8,102 rows, each corresponding to a gene, and 84 columns,
each corresponding to a sample.  PAL first selected the subset of
genes whose expression varied by at least 4-fold from the median
of the samples, in at least three of the samples tested. This
filtering process left the set $G1$ of 1753 genes, each of which
is represented by 84 expression values. In the final expression
matrix PAL split the data into two submatrices; one of tissues and
one of cell lines. The two submatrices were, separately, median
polished (the rows and columns were iteratively adjusted to have
median 0) before being rejoined into a single matrix. The
expression matrix was two-way clustered; clustering the genes on
the basis of the 84 samples [operation $G1(S)$], and clustering
the 65 tumors using all 1753 genes [$S1(G1)$]. Since  $S1(G1)$ did
not yield any meaningful partition, PAL concluded that the 1753
genes were not an optimal set to classify the tumors, and they
selected a subset $G^{(int)}$ of 496 "intrinsic" genes in the
following way. They calculated for each gene an index that
measures the variation of it's expression between different tumors
versus between paired samples from the same tumor. They ranked all
8102 genes according to this index, and chose the 496 top scorers.
They argued that the expression levels of the top scorers on this
list represent inherent properties of the tumors themselves rather
than just differences between different samplings. From this point
on they used the $496 \times 65$ expression level matrix  to
cluster the genes of $G^{(int)}$ and the tumors $S1$. This data is
publicly available at the Stanford website (see PAL).


The second study of breast cancer, by SAL, characterized gene
expression profiles of 85 tissue samples representing 84
individuals. 78 of these were breast carcinomas (71 ductal, five
lobular, and two ductal carcinomas in situ, obtained from 77
different individuals; two tumors were from one individual,
diagnosed at different times) 3 were fibroadenomas and 4 were
normal breast tissue samples were also included; three of these
were pooled normal breast samples from multiple individuals
(CLONTECH). These 85 samples included 40 tumors that were
previously analyzed and described by PAL. Fifty-one of the
patients were part of a prospective study on locally advanced
breast cancer (T3/T4 and/or N2 tumors) treated with doxorubicin
monotherapy before surgery followed by adjuvant tamoxifen in the
case of positive ER and/or progesterone receptor (PgR) status
(Geisler \al, 2001). All but three patients were treated with
tamoxifen. ER and PgR status was determined by using
ligand-binding assays, and mutation analysis of the TP53 gene was
performed as described in Geisler \al. The cDNA microarrays used
in this study were from several different print runs that all
contained the same core set of 8,102 genes. In total, the 85
microarray experiments were carried out by using six different
batches of microarrays and three different batches of common
reference, each independently produced. SAL performed cluster
analysis on two subsets of genes. One subset, of 456 cDNA clones
(427 unique genes), was selected from the 496 "intrinsic" gene
list, previously described by PAL. The second subset consisted of
264 cDNA clones, that exhibit high correlation with patient
survival, were selected from the set $G1$ of 1753 genes.
Clustering analysis and patient classifications were based on the
total set of 78 malignant breast tumors. Survival analysis was
based on 49 patients with locally advanced tumors and no distant
metastases (two of the 51 patients from this prospective study
were retrospectively recorded to have a minor lung deposit and a
liver metastasis, respectively) that were treated with neoadjuvant
chemotherapy and adjuvant tamoxifen
(Geisler \al, 2001)\\
{\bf Expression data - colon cancer.} In addition, we studied a
data set on colon cancer, previously published by Notterman et al.
The data set contains 22 tumor samples; 18 carcinoma and 4
adenoma, and their paired normal samples. The experiments with
carcinoma and paired normal tissue were performed with the Human
6500 GeneChip Set (Affymetrix), and the experiments with the
adenomas and their paired normal tissue were performed with the
Human 6800 GeneChip Set (Affymetrix). First, following Notterman
et.al, we created a composite database that included only
accession numbers represented on both GeneChip versions. Values
lower than 1 were adjusted to 1. Prior to application of CTWC, we
filtered the data using a filtering operation very close to that
used by Notterman et.al, remaining with 1592 genes. Data from the
two different chips were brought to the same average expression
level. The data was then log-transformed, centered about the mean
and normalized. Second, we studied the 18 paired carcinoma samples
separately. Of the ~6600 cDNAs and ESTs represented on the array,
only genes for which the standard deviation of their
log-transformed  expression values was greater than 1, were
selected. After this filtering process we remain with 768 genes.
These values were centered and normalized, prior to application of
the CTWC algorithm. The samples were labeled according to
additional information about the histological characteristics of
the tumor samples, the estimated percentage of contamination with
non-tumor cells, the presence of mutations in the p53 gene, the
clinical disease stage and the mRNA extraction protocol that has
been used.

\section*{ALGORITHM}
Since both SPC and CTWC have been described in detail elsewhere,
we present here only brief, albeit self contained reviews of the
procedures.\\

\noindent
{\bf SUPERPARAMAGNETIC CLUSTERING - SPC} \\

The idea behind this algorithm is rooted in the physics and phase
transitions of disordered magnets (Blatt \al, 1996; for a detailed
description see Blatt \al, 1997). The four-step procedure
presented here uses terminology of graph theory, which is more
familiar to computer scientists.

{\bf Step 1: Weighted graph.} $N$ data points are associated with
"positions" $\X_i$ in a $D$-dimensional space; they constitute $N$
nodes of a graph. Each node is connected by an edge to its
neighbors. We identify the neighbors $j$ of each node $i$ on the
basis of the distances $d_{ij} = \vert \X_i - \X_j \vert $
\footnote{Normally Euclidean distances are used.}; the two points
are neighbors if $j$ is one of the $K$ closest neighbors of $i$,
and vice versa\footnote{$K$ is a parameter of the algorithm - for
genes  we use $10 \leq K \leq 20$. By superimposing the minimal
spanning tree, we ensure that all vertices belong to a single
connected component of the graph.}. To each edge $(ij)$ we assign
a weight $J_{ij}= f(d_{ij})$ where $f(x)$ is a decreasing
function\footnote{We use $f(x) = (1/\sqrt{2\pi} a) {\rm exp}[
-x^2/2a^2 ]$ ($a$ is the average of $d_{ij}$).} of $x$.

{\bf Step 2: Cost function for graph partitions:} To characterize
a partition of the graph, we assign to every vertex $i$ an integer
label (a Potts spin variable in Physics terminology),
$S_i=1,2,...q$ \footnote{In many of the applications we tried,
Potts spins with $q=20$ states were used. $q$ has nothing to do
with the number of clusters determined by the algorithm - see
below.}. Any particular assignment of labels, $\{S_1,S_2,...S_N
\}$, is represented as a "configuration" $\{ S \}$. $S_i=S_j$
indicates that in the partition $\{ S \}$, nodes $i$ and $j$
belong to the same component, whereas $S_i \neq S_j$ means that
they are in different components. The cost function
 \beq {\cal
H}(\{ S \} )= \sum_{ <i,j>} J_{ij} \; \left ( 1 - \delta_{ S_i,
S_j} \right ) \label{eq:Hamiltonian}
\eeq
places a high penalty $J_{ij}$ on assigning two similar nodes
$i,j$ (i.e. with small $d_{ij}$), to different components. The
lowest cost, ${\cal H}(\{ S \} )=0$ is obtained when all data
points are in the same group; the highest cost is reached if no
point is in the same group as any of it's neighbors. Hence the
value of ${\cal H}(\{ S \} )$ reflects the resolution at which the
partition $\{ S \}$ views the data.

{\bf Step 3: Ensemble of partitions.} Rather than choosing any
particular partition (say by minimizing the cost function), we
consider all configurations $\{S \}$ that have (nearly) the same
value of ${\cal H}(\{ S \})=E $; to each of these we give the {\it
same statistical weight}, whereas all $\{ S' \}$ that correspond
to different resolutions (and hence ${\cal H}($\{ S' \}$) \neq E$)
get vanishing probability. The resulting ensemble of partitions
$\{ S \}$ is the microcanonical ensemble of Statistical Mechanics.
For each value of $E$ one can sample this ensemble and measure
average values of any quantity of interest (see below). It is,
however, more convenient to use for such measurements the
canonical ensemble, in which rather than fixing the value of $\cal
H$, we control the value of its ensemble average by a Lagrange
multiplier, $1/T$. In the resulting statistical ensemble of
partitions  each $\{ S \}$) appears with the statistical
(Boltzmann) weight
\beq P(\{ S \}) \propto e^{-{\cal H}(\{ S\})/T}
\label{eq:Boltz}
\eeq
At $T=0$ only groupings with $E=0$ have non-vanishing weight; at
$T=\infty$ all partitions have equal weight. For a sequence of
values of the {\it temperature T} we calculate, by Monte Carlo
simulation, the equilibrium average $\Taver{A}$ of several
quantities $A$ of interest, such as the magnetization,
susceptibility and correlation of neighbor spins:
\begin{equation}
  \label{eq:Gij}
   G_{ij} = \Taver{ \delta_{S_{i},S_{j}} } \; ,
\end{equation}
$G_{ij}$ is the probability to find, at the resolution set by $T$,
data points $i,j$ in the same group. By the relation to granular
ferromagnets we expect that the distribution of $G_{ij}$ is
bimodal; if both spins belong to the same {\it ordered} grain
(cluster), their correlation is close to 1; if they belong to two
clusters that are not relatively ordered, the correlation is close
to $1/q$.

{\bf Step 4: Identifying clusters.} To produce "hard" clusters on
the basis of the $G_{ij}$,
we construct a new graph, in a three-step procedure.\\
1. Build the clusters' ``core'' by thresholding $G_{ij}$. For
every pair of neighbors $i$ and $j$, check whether $G_{ij}> \theta
=0.5$; if true, set a
"link" between $i,j$. Because of the bimodality of the distribution of $G_{ij}$
the decision to link $i,j$ depends very weakly on the value of $\theta$. \\
2. Capture points lying on the periphery of the clusters by
linking each point $i$ to its neighbor $j$ of maximal correlation
$G_{ij}$.\\
3. Data clusters are identified as the linked components of the
graphs obtained in steps 1,2.\\

At $T=0$  this procedure generates a single cluster of all $N$
points. At $T=\infty$ we have $N$ independent spins, and the
procedure yields $N$ clusters, with a single point in each. Hence
as $T$ increases, we generate a dendrogram of clusters of
decreasing sizes.

This algorithm has several attractive features (Blatt \al, 1997),
one of these is the ability to identify stable (and statistically
significant) clusters, which makes SPC most suitable to be used
within the framework of CTWC. Furthermore, it allows a
quantitative estimation for the P-value of a clustering operation,
by clustering repeatedly randomized data and checking the fraction
of instances in which stable clusters (i.e. as stable as those
obtained for non random data) appeared. We identify stable
clusters as follows. As we heat the system up, we record for every
cluster two temperatures: $T_1$, at which it is "born" (splits
from it's parent cluster) and $T_2$, at which it "dies" (splits
into siblings). The ratio  $R= T_2 / T_1$ is a measure of a
cluster's stability.

SPC was used in a variety of contexts, ranging from computer
vision (Domany \al, 1999) to speech recognition (Blatt \al, 1997).
Its first direct application to gene expression data has been
(Getz \al, 2000) for analysis of the temporal dependence of the
expression levels in a synchronized yeast culture (Eisen \al,
1998), identifying gene clusters whose variation reflects the cell
cycle.

Subsequently, SPC was used to identify primary targets of p53
(Kannan \al, 2001) and p73 (Fontemaggi \al, 2002).\\

\noindent
{\bf COUPLED TWO WAY CLUSTERING - CTWC}\\

The main motivation for introducing CTWC (Getz \al, 2000) was to
{\it increase the signal to noise ratio} of the expression data.
The method is designed to overcome two different kinds of "noise".
The first was mentioned above; say only a small subset of $N_r$
genes participate in a biological process of interest, associated
with a particular disease $A$. In this case we expect these $N_r$
genes to have correlated expressions over subjects with disease
$A$. This correlation could, in principle, identify the diseased
subjects as "close" in expression space - but, in fact, for $N_r
<< \vert G1 \vert$ the non-participating $\vert G1 \vert - N_r$
genes completely mask the effect of the relevant ones on the
distance between two diseased subjects. Hence as far as the
process of interest is concerned, the non-participating $\vert G1
\vert - N_r$ genes contribute nothing but noise, that masks the
signal of the $N_r$ relevant ones. CTWC eliminates this noise by
discarding the irrelevant genes.

The second noise-reducing feature of CTWC is that it uses the expression levels
of a a set of genes, rather than one gene at a time. Thereby intrinsic noise
in the expression averages out.

CTWC is an iterative process, whose starting point is the standard
two way clustering mentioned above, i.e. the clustering operations
$S1(G1)$ and $G1(S1)$. We keep two registers - one for stable gene
clusters and one for stable sample clusters. Initially we place
$G1$ in the first and $S1$ in the second. From  $S1(G1)$ and
$G1(S1)$ we identify stable clusters of samples and genes,
respectively, i.e. those for which the SPC stability index $R$
exceeds a critical value and whose size is not too small. Stable
gene clusters are denoted as $GI$ with $I=2,3,...$ and stable
sample clusters as $SJ, J=2,3,...$ In the next iteration we use
every gene cluster $GI$ (including $I=1$) as the feature set, to
characterize and cluster every sample set $SJ$. These operations
are denoted by $SJ(GI)$ ($S1(G1)$ was already performed). In
effect, we use every stable gene cluster as a possible "relevant
gene set"; the submatrices defined by $SJ$ and $GI$ are the ones
we study. Similarly, all the clustering operations of the form
$GI(SJ)$ are also carried out. In all clustering operations we
check for the emergence of partitions into stable clusters, of
genes and samples. If we obtain a new stable cluster, we add it to
our registers and record its members, as well as the clustering
operation that gave rise to it. If a certain clustering operation
did not give rise to new significant partitions, we move down the
list of gene and sample clusters to the next pair.

This heuristic identification of relevant gene sets and
submatrices is nothing but an exhaustive search among the stable
clusters that were generated. The number of these, emerging from
$G1(S1)$, is a few tens, whereas $S1(G1)$  usually generates only
a few stable sample clusters. Hence the next stage typically
involves less than a hundred clustering operations. These
iterative steps stop when no new stable clusters beyond a preset
minimal size are generated, which usually happens after the first
or second level of the process.

In a typical analysis we generate between 10 and 100 interesting
partitions, which are searched for biologically or clinically
interesting findings, on the basis of the genes that gave rise to
the partition and on the basis of available clinical labels of the
samples. It is important to note that these labels are used {\it a
posteriori}, after the clustering has taken place, to interpret
and evaluate the results.

\section*{RESULTS}
Lists of the genes that constitute each of the clusters $GI$
mentioned below are given in the supplementary information.\\

\noindent
{\bf BREAST CANCER - PAL} \\

We posed the following questions:\\
1. Do our methods of analysis reproduce the results obtained by
PAL?\\
2. Can we make observations that seem to be of interest and were
not reported by PAL?\\
As to the first question - CTWC reproduced all the main findings
of PAL directly, starting from the entire set $G1$ of 1753 genes,
without filtering them to the "intrinsic set". Second, we found
new tumor classifications that were not mentioned
by PAL.\\
{\bf Reproducing the results of PAL:}

PAL used lower case letters to identify gene clusters, and colors
for samples (see their Figs. 1 and 3). We use below their notation
when comparisons are made.

$G1(S)$: Following PAL, we used the same feature set, $S$, of all
samples and cell lines, to cluster $G1$, the full set of 1753
genes. Since we also used the same normalization, this operation
provides a direct comparison of Average Linkage (the clustering
method used by PAL) and SPC. All the gene clusters that were
marked as interesting by PAL, were also found by our clustering
operation (Kela 2001).

$S(G1)$: Next, we clustered (separately) the cell lines and the
tumors, using all 1753 genes. Since our normalization here differs
from that of PAL, we cannot compare directly our results. However,
in agreement with PAL, we also did not find any meaningful
partitions of the tumors, $S1$, from this operation, leading to
the same conclusion as reached by PAL: namely, that $G1$ is not
suitable to classify the tumors and we should characterize them
using different subsets of genes. From here on CTWC deviates from
the procedure of PAL, who selected their "intrinsic set" of 496
genes in a way that (a) necessitates having paired samples from
the same patients ({\it before} and {\it after} chemotherapy), and
(b) assumes that only genes that meet their criteria (similarity
of matched samples) are to be used. CTWC, on the other hand, is an
automated process, performing operations $S1(GI)$, i.e. clustering
the tumors $S1$ using different stable gene clusters $GI$, one at
a time. Clustering the 65 samples on the basis of these small
subsets of genes, one at a time, enabled us to identify the
subclasses of tumors that PAL found using their intrinsic set.

$S1(G4)$: cluster $G4$ (that was obtained by the G1(S) clustering
process) has 10 genes - it is our homologue of cluster j of PAL
(see their Fig. 1). The operation $S1(G4)$ generates a stable
sample cluster which is quite similar to the ER+/luminal-like
(blue) cluster of PAL (see their Fig.~3); its members have high
expression levels of $G4$. $S1(G4)$ identifies also PAL's
basal-like (yellow) group, characterized by low expression levels
of the $G4$ genes.

$S1(G46),S1(G9)$: $G46$ is a cluster of 33 genes that are part of
the proliferation cluster found by PAL. The operation $S1(G46)$
produces a good homologue of their normal-like (green) cluster.
Members of this group show low expression levels of $G46$ genes.
The normal-like samples are also identified in the operation
$S1(G9)$: the 13 genes of $G9$ are a subgroup of cluster g of PAL.
Normal-like tissues have high expression levels of the $G9$ genes.

$S1(G21)$: This operation separates  the Erb-B2+ (red) cluster
from the other samples. $G21$ is homologous to gene cluster d from
Fig. 3 of PAL; it's expression is high in the Erb-B2+  tumors.
\\

{\bf New observations (beyond PAL):}\\
Of several new findings (Kela 2001) we chose to highlight here one
that bears on an issue that has been considered important by PAL:
that of separating the ER+ and ER- tumors on the basis of their
expression levels. We present two such classifiers, which
demonstrate two different advantages of CTWC. The first classifier
{\it could have been} discovered by PAL, since it is based on
genes that {\it do belong to PAL's intrinsic set}, but their
effect is masked by the large number of the 496 "intrinsic" genes;
to see it, one has to zero in on a small subset, as is done by
CTWC. The second classifier {\it could not have been discovered by
PAL's analysis} since it is based on genes that are {\it not
included in their intrinsic set}.

$S1(G4):$ The cluster $G4$ (10 genes) was described above - it is
practically identical to cluster j from Fig.~1 of PAL and to
cluster c of their Fig.~3. It contains the estrogen receptor and
three other transcription factors (see supplementary information
of PAL) related to the estrogen receptor pathway. The operation
$S1(G4)$ generated the dendrogram presented in Fig.~1A. The
variation in the expression levels of the $G4$ genes correlates
well with the direct clinical measurements of the ER protein
levels in the tumors (supplementary information of PAL).

\begin{figure}
    \centerline{
       \psfig{figure=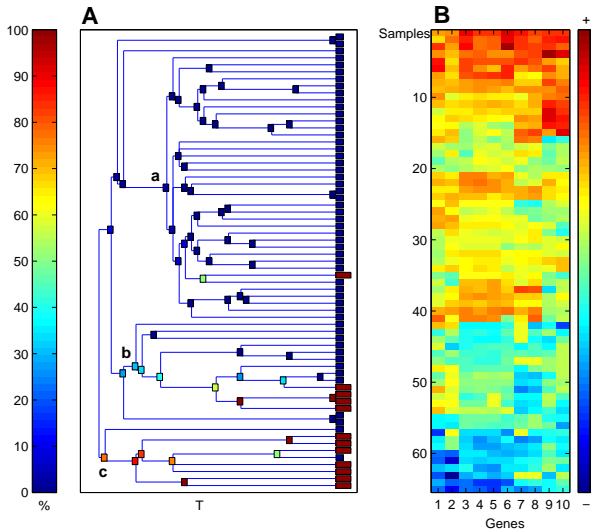,width=8.0cm}
     }
\caption{$S1(G4)$: clustering 65 tumors using the expression
levels of gene cluster $G4$. A. The boxes in the dendrogram
represent clusters; they are colored according to their percentage
of ER- tumors (see color bar on left). B. Clusters {\bf a,b,c} are
characterized, respectively,  by high, intermediate and low
expression levels (see color bar on right).}
\label{fig:s1(g4)}
\end{figure}

In the dendrogram Fig. 1A the boxes representing sample clusters
were colored according to the percentage of ER- samples, ranging
from red (100\%) to blue (0\%). In Fig. 1B the samples were
ordered according to the dendrogram, and the colors represent the
expression levels of the 10 genes. SPC generated 3 main branches
(clusters); the upper - {\bf a} with highest expression values,
{\bf b} - intermediate and the lowest - {\bf c}. Cluster {\bf a},
the biggest (41 samples), contains all but two of the tumors of
the luminal-like (blue) cluster of PAL (see their Fig.~3). More
interestingly, clusters {\bf a} and {\bf b}, contain 45 out of 48
of the ER+ tumors (see blue leaves). Cluster {\bf c} is rich (7
out of 11) in ER- tumors. Designating any sample that {\it does
not} belong to {\bf c} as $ER+$, we get our best classifier, with
efficiency $E=45/48=0.94$ and purity $P=0.83$. The corresponding
numbers obtained by PAL (for their "luminal-like" cluster) were
$E=0.66$ and $P=0.89$.

$S1(G30):$ $G30$ is a cluster of 15 genes, related to cell cycle
proliferation. Only one of the 15 were included in PAL's intrinsic
set. Clustering the 65 tumors using the expression levels of these
genes generated the dendrogram presented in Fig.~5A (see
supplementary information). The boxes that represent sample
clusters are colored according to their relative content of ER-
samples. The dendrogram exhibits a clear partition of the tumors
into clusters {\bf a} with high expression levels of the $G30$
genes and {\bf c} with intermediate expression levels, as seen in
Fig.~5B. Cluster {\bf c} contains 44 tumors, 38 of which were
classified as ER+, 3 as ER- and 3 unknown. Hence this cluster
captured the ER+ group with efficiency of $E= 38/48=0.79$ and
purity $P=38/44 = 0.86$. Cluster {\bf a} contains a high
proportion of ER- tumors; its sub-cluster {\bf b} consists of 5
'special' ER+ tumors that have relatively high expression levels
of the $G30$ genes.
\\

\noindent
{\bf BREAST CANCER - SAL}\\

Again we have two kinds of observations; those made using genes
that were not included by SAL in their intrinsic set, and hence
could not have been found by them, and observations made using
genes that were included in the previous analysis.

Since there is considerable overlap between the samples of PAL and
SAL, we did not repeat our attempt to reproduce all their
findings. We did, however, study some aspects related to the
clinical labels, that were the main additional feature of the SAL
data. We emphasize here our findings concerning survival and p53
status. We found correlations between expression levels of several
gene clusters and survival, and that the expression levels of
these genes is also a predictor of p53 mutation status. We also
present a very clear partition of the patients into two groups,
for which we do not yet have any clinical interpretation.

$S1(G10):$ cluster $G10$ contains 15 genes that are related to the
ER pathway, including 5 of the 10 members of $G4$ mentioned in our
analysis of PAL, (such as GATA-binding protein 3). Clustering the
85 samples ($S1$) using $G10$, generates the dendrogram presented
in Fig. 6A (supplementary information). The boxes that represent
sample clusters are colored according to the median value of the
survival of the patients contained in each cluster, ranging from
red (median survival of 100 months) to dark blue (4 months).
Similarly to the results shown in Fig.~1, the variation in the
expression levels of the $G10$ genes correlates well with the
direct clinical measurements of the ER protein levels in the
tumors. The dendrogram of Fig.~6A exhibits two main clusters; {\bf
a} contains most of the ER+ tumors, that exhibit higher expression
levels of the $G10$ genes, as seen in Fig. 6B, and {\bf b}, which
contains mainly ER- tumors that exhibit low expression levels of
the $G10$ genes.

Analyzing the correlation with the p53 status, wild type (wt) vs
mutant, and with the survival parameter we get similar results as
were obtained by SAL. They showed that the basal-like samples,
corresponding to our cluster {\bf b}, come from patients with the
shortest survival times and a high frequency of p53 mutations. Two
of the 17 members of cluster {\bf b} survived for 41 months and
all the others - for less than 26 months. The correlation
coefficient between survival and the average expression levels of
the $G10$ genes is {\bf 0.47}. The Wilcoxon rank-sum test (WRST)
indicated that to the distributions of survival times in cluster
{\bf b} to the rest of the patients are significantly different
(P-value = $3.7 10^{-4}$); patients that exhibit low expression
levels of the $G10$ genes have short survival.

To indicate the p53 status, we placed a color bar next to the
leaves of the dendrogram, on which the patients with mutant p53
are labeled red and the p53 wt - blue. Patients with unknown p53
status were labeled white. Note that the 17 patients of cluster
{\bf b} exhibit low expression levels of the $G10$ genes. Ten of
these 17 are p53 mutant, 5 have unknown labels and only two are
wt. Hence low expression levels of the $G10$ genes seem to go
along with a mutated p53. The correlation coefficient of the
average expression levels of $G10$ with p53 status is {\bf 0.4};
in particular, low expression is a good predictor of mutant p53.
To substantiate the last statement, we compared the distributions
(using WRST) of the median expression levels of patients with
mutant p53 to wt. We found that the two distributions are
significantly different (P-value=$1.2 10^{-4}$); the wt p53
patients exhibit high and the mutant p53 exhibit lower expression
levels of the $G10$ genes.

$S1(G33):$ Cluster $G33$ contains 36 genes, related to cell
proliferation, which include 10 out of the 15 members of cluster
$G30$ found by CTWC in our analysis of the PAL data. Clustering
the 85 samples using the expression levels of these genes
generated the dendrogram presented in Fig.~2A. The boxes are
colored similarly to Fig. 6A; according to the median survival (in
months), of the patients that belong to each cluster. The $G30$
genes partition the samples into 3 main clusters, {\bf a, b} and
{\bf c}, as shown in the dendrogram. The corresponding $G33$
expression levels, as seen in Fig.~2B, are high, intermediate and
low, respectively. The average expression level of the $G30$ genes
is inversely correlated with survival (correlation coefficient
{\bf -0.24}). Cluster {\bf a} contains patients with high
expression and short survival; only one of its 21 members survived
beyond 43 months, whereas clusters {\bf b} and {\bf c} contain
long (up to 100 months) as well as short survival. Comparison of
the distributions of the survival times of the patients in cluster
{\bf a} to those in clusters {\bf b} and {\bf c} indicates that
there is a significant difference (P-value = 0.0016).

As to p53 status, we note that among the 21 patients in cluster
{\bf a}, 13 were mutant p53 and 4 had unknown status. Cluster {\bf
c}, of low expression levels, contains only two mutant p53
patients (out of 16 members of the cluster). The correlation
coefficient between the average expression levels of $G33$ genes
and p53 status is {\bf -0.4}. Hence high expression levels of
these genes is a good predictor for mutant p53, whereas low
expression predicts wt p53. Comparison of the distributions of the
median expression levels between the p53-mutant and  the p53-wt
patients yields significantly different distributions
(P-value=$4.510^{-5}$).

\begin{figure}
    \centerline{
       \psfig{figure=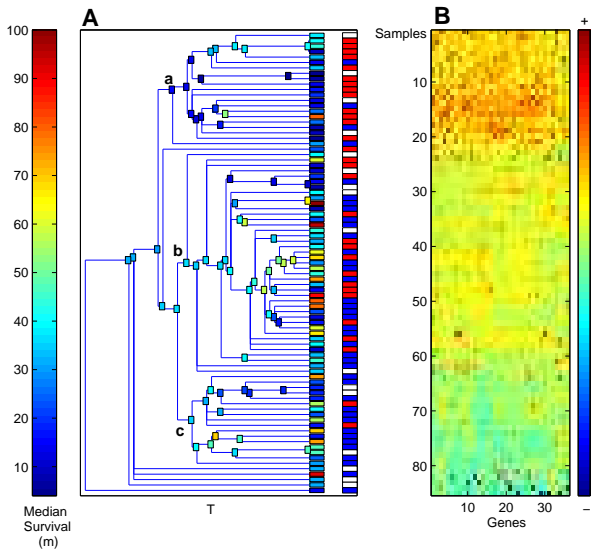,width=8.0cm}
     }
\caption{$S1(G33)$: The boxes in the dendrogram A represent sample
clusters that are colored according to the median value of the
survival of the patients contained in each cluster, ranging from
dark red (median survival of 100 months) to blue (median - 4
months) - see left color bar. B. The clusters {\bf a,b,} and {\bf
c} exhibit high, intermediate and low expression levels (see color
bar at right). The central color bar represents p53 status: red -
mutant, blue - wt and white - unknown. Members of {\bf a} are
characterized by high expression, low survival and mutant p53.}
\label{fig:s1(g33)}
\end{figure}

$S1(G36):$ cluster $G36$ contains genes that are related to
apoptosis suppression (e.g. bcl-2) and cell growth inhibition
(e.g. INK4C – cyclin-dependent kinase inhibitor 2c).  Using the
expression levels of this set of genes to cluster the 85 samples,
we generate the dendrogram presented in Fig.~3A. The boxes are
colored similarly to Fig.~2A, according to the median survival of
the patients in each cluster. The dendrogram exhibits partition of
the samples into two very distinct clusters; {\bf a} contains
patients with high expression levels and {\bf b} - patients with
low. We found no correlation between membership in either of these
clusters and any of the clinical labels that were reported by SAL.
However, the clarity of the partition calls for further
investigation of the two groups of patients, which may reveal some
so far unknown role played
by the genes of $G36$ in breast cancer.\\

\begin{figure}
    \centerline{
       \psfig{figure=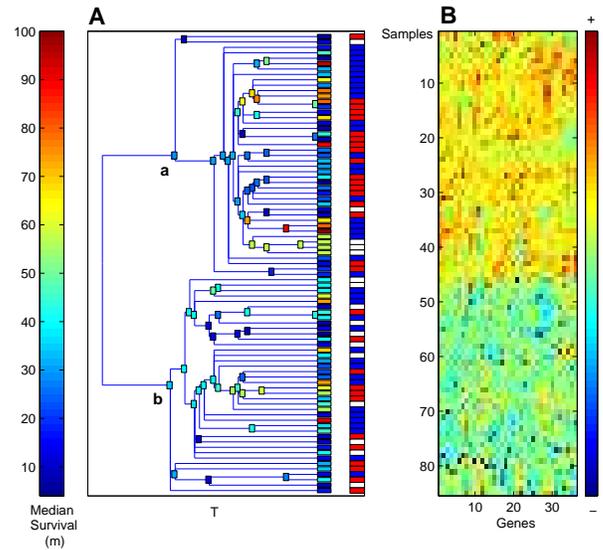,width=8.0cm}
     }
\caption{$S1(G36)$: A. The genes of $G36$ gave rise to a very
clear partition of the samples to high (cluster {\bf a}) and low
expression levels {\bf b}. No clinical interpretation of this
partition has been found yet.} \label{fig:s1(g36)}
\end{figure}

\noindent
{\bf COLON CANCER}\\

We applied CTWC on the colon data set of Notterman \al, containing
18 paired carcinoma and 4 paired adenoma samples. We refer to the
set of all 44 samples as S and to the 36 paired carcinoma samples
as S1. We present gene clusters which differentiate the samples
according to the known normal/tumor classification, previously
shown by Notterman \al. Furthermore, we show the advantage of CTWC
in mining new partitions  which have not been found using other
clustering methods and may contain
relevant biological information.\\
{\bf Tumor - Normal separation}

$S(G8):$ $G8$ contains 55 genes, which show high expression levels
in the normal samples compared to the adenoma and carcinoma.
Several genes within this cluster are known to be repressed in
colorectal neoplasms;  for example, guanilyn and DRA
(down-regulated in adenoma). Some of these genes were previously
mentioned by Notterman \al Clustering the 44 samples, using the
expression levels of  $G8$, generated the dendrogram shown in
Fig.~4A.

\begin{figure}
    \centerline{
       \psfig{figure=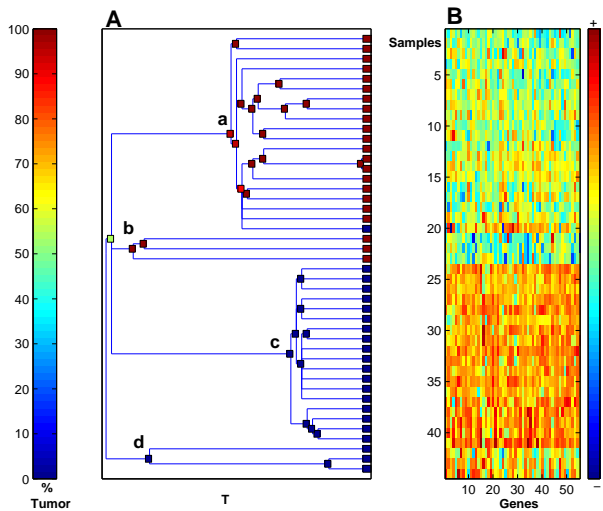,width=8.0cm}
     }
\caption{$S(G8)$: A clear separation of the tumor carcinoma and
adenoma samples from the normal samples, using the $G8$ group of
genes. (A) The boxes are colored according to the percentage of
the tumor samples. (B) The expression level matrix of $S1(G8)$.
Rows correspond to all the samples and the columns correspond to
the genes of cluster $G8$. The matrix shows relatively high
expression levels of the $G8$ genes in the normal samples compared
to the tumor samples.} \label{fig:s1(g8)}
\end{figure}

The dendrogram exhibits a clear separation into two large clusters
({\bf a} and {\bf b}) and two small ones ({\bf c} and {\bf d}).
Clusters {\bf c} and {\bf d} contain all the normal samples (both
carcinoma and adenoma), {\bf a} - the tumor carcinoma samples and
{\bf b} - the tumor adenoma samples. The colors (see bar on the
right hand side of the expression matrix - see reordered data)
represent the expression levels of the genes in $G8$, with red
(blue) denoting high (low) values.

$S1(G25):$ The data set we analyzed next contains the 18 carcinoma
and their paired normal samples, $S1$. The group $G25$ contains 51
genes, some of which are known to be over expressed in carcinoma
and are found to be related to colon cancer or other forms of
neoplasma e.g, myc, matrilysin, GRO-$\gamma$ (see Notteman \al,
2001), and additional genes which may very well be related to
colon cancer. Clustering the 36 samples of $S1$, using the
expression levels of the gene cluster $G25$, gave rise to a clear
partition of the samples into two clusters; one of normal samples
({\bf a}), and the other of tumor samples ({\bf b}), with
relatively high expression levels of the $G25$ genes in the tumor
cluster (see Fig. 7,
supplementary information).\\
{\bf New observations (protocols A,B):} \\
$S1(G3):$ Two experimental protocols that were used; 16 RNA
samples (paired samples 3-6,8-10,11) were extracted using a method
that isolates mRNA prior to reverse transcription ('protocol A'),
and the other 20 samples (paired samples 12,27,28-29,32-35,39-40)
were prepared by extracting total RNA from the cells ('protocol
B'). Clustering the 36 carcinoma samples, using the expression
levels of the 27 genes of cluster $G3$, exhibits a clear partition
of the samples into two clusters (see Fig~8A, supplementary
information). Cluster {\bf b} contains 20 tissues of protocol B,
and cluster {\bf a} contains 14 tissues of protocol A.  This
separation has two mistakes; both samples of patient 9 were
labeled A and appear in
the cluster of protocol B.\\
{\bf New observations (unknown interpretation):}

$S10(G24), S10(G7), S10(G12):$ Clustering only the 18 carcinoma
samples ($S10$, obtained in a previous CTWC iteration) on the
basis of their expression over different sets of genes, revealed
the following partitions:

The clustering operation $S10(G24)$  generated a clear separation
of the tumor samples into 2 clusters. Samples 33,34,35,40 are
clustered together in {\bf b}, and show high expression levels of
the $G24$ genes (Fig.~9, supplementary information).

The operation $S10(G7)$ separated tumor samples 27,32,33,40 from
the other 14; the small group has low expression levels of the
$G7$ genes (Fig.~10, supplementary information) .

$S10(G12)$ clustered tumor samples 33,34,35,12,40 together
(cluster {\bf b} in Fig.~11, supplementary information); the
expression levels of the $G12$ genes are high in these 5 samples.
Hence we discovered that tumor samples 33,40 and 35 were
repeatedly separated from the remaining tumors, which implies that
these patients may share some  common characteristics, perhaps
representing a true biological meaning. However, due to lack of
additional information about the patients we were unable to
determine the biological origin of this separation. \vspace{0.3cm}

\noindent   {\bf DISCUSSION AND CONCLUSION}

We described the  {\it Coupled Two Way Clustering} method and
demonstrated it's ability to extract useful information from
breast cancer and colon cancer data. For both data sets we
reproduced the findings of previous analyses and discovered new
structure of biological significance, demonstrating the advantages
of CTWC compared to standard clustering techniques.

The central strategy of CTWC is to cluster the samples on the
basis of their expression levels over small, correlated sets of
genes, and vice versa. The relevant sets of genes and samples are
found by using, one at a time, stable  clusters of genes (or
samples), that were identified in preceding iterations of the
algorithm. Whenever such a clustering operation generates new,
statistically significant partitions of the clustered objects, the
result is recorded, to be used in further iterations and to be
scanned for possible biological or clinical interpretation.

Perou \al~also reached the conclusion that performing an "all
against all" analysis does not reveal the effects of relatively
small groups of relevant genes. They were able to produce
significant findings only after reduction of the genes used to a
smaller number. The smaller "intrinsic set" was identified using a
particular guiding principle, one that can be used only when there
are at least two samples from each of several patients.
Furthermore, the selection criteria used exclude genes that,
according to our findings, do contain important information.

CTWC does not only generate the important partitions of the
samples; it also identifies small groups of genes that are
responsible for the separation of different classes. For both
breast and colon cancer we found partitions that have no clear
interpretation at the moment, a fact that demonstrates the
strength of unsupervised approaches such as clustering;
unsuspected structure buried in the data can be revealed.

\vspace{0.3cm}

\noindent   {\bf ACKNOWLEDGEMENTS}

\noindent This research was partially supported by grants from the
Germany - Israel Science Foundation (GIF), the Israel Academy of
Sciences (ISF) and the Ridgefield Foundation. We thank D. Botstein
for directing us to the two papers of the Stanford group on breast
cancer (PAL and SAL).

\vspace{0.3cm}

\noindent {\bf REFERENCES}
{\small

\noindent Alon,U., Barkai,N., Notterman,D.A., Gish,K., Ybarra,S.,
Mack,D. and Levine,A.J. (1999) Broad patterns of gene expression
revealed by clustering analysis of tumor and normal colon tissues
probed by oligonucleotide arrays, {\it Proc. Natl. Acad. Sci.
USA}, {\bf 96}, 6745--6750.

\noindent Blatt,M., Wiseman,S. and Domany,E. (1996)
Superparamagnetic clustering of data, {\it Phys. Rev. Lett.}, {\bf
76}, 3251--3254.

\noindent Blatt,M., Wiseman,S. and Domany,E. (1997) {\it Neural
Comp.} Data Clustering Using a Model Granular Magnet, {\bf 9},
1805--1842.

\noindent Califano,A., Stolovitsky,G. and Tu,Y. (2000) Analysis of
Gene Expression Microarrays for Phenotype Classification, {\it
Proc. Int. Conf. Intell. Syst. Mol. Biol.}, {\bf 8}, 75--85.

\noindent Cheng,Y. and Church,G.M. (2000) Biclustering of
Expression Data, {\it Proc. Int. Conf. Intell. Syst. Mol. Biol.},
{\bf 8}, 93--103.

\noindent Domany,E., Blatt,M., Gdalyahu,Y. and Weinshall,D. (1999)
Super-paramagnetic clustering of data: application to computer
vision, {\it Comp. Phys. Comm.}, 121--122.

\noindent Eisen,M.B., Spellman,P.T., Brown,P.O. and Botstein,D.
(1998) Cluster analysis and display of genome-wide expression
patterns, {\it Proc. Natl. Acad. Sci. USA}, {\bf 95},
14863--14868.

\noindent Fontemaggi,G., Kela,I., Amariglio, N., Rechavi, G.,
Krishnamurthy,J., Strano, S., Sacchi, A.,Givol, D. and Blandino,
G. (2002) Identification of direct p73 target genes combining DNA
microarray and chromatin immunoprecipitation analyses; Comparison
with p53 targets, {\it (unpublished)}.

\noindent Geisler,S., Lonning,P.E., Aas,T., Johnsen,H., Fluge,O.,
Haugen,D.F., Lillehaug,J.R., Akslen,L.A. and Borresen-Dale,A.L.
(2001) Influence of TP53 gene alterations and c-erbB-2 expression
on the response to treatment with doxorubicin in locally advanced
breast cancer, {\it Cancer Res.}, {\bf 6}, 2505-2512.

\noindent Getz,G., Levine,E., Domany,E. and Zhang,M.Q. (2000)
Super-paramagnetic clustering of yeast gene expression profiles,
{\it Physica A}, {\bf 279}, 457--464.

\noindent Getz,G., Levine,E. and Domany,E. (2000) Coupled two-way
clustering analysis of gene microarray data, {\it Proc. Natl.
Acad. Sci. USA}, {\bf 97}, 12079--12084.

\noindent Getz,G. and Domany,E. (2002) Coupled Two-Way Clustering
Server, (this issue).

\noindent Godard,S. {\it et al,unpublished}

\noindent Golub,T.R., Slonim,D.K., Tamayo,P., Huard,C.,
Gaasenbeek,M., Mesirov,J.P., Coller,H., Loh,M.L., Downing,J.R.,
Caligiuri,M.A., Bloomfield,C.D. and Lander,E.S. (1999) Molecular
classification of cancer: class discovery and class prediction by
gene expression monitoring., {\it Science}, {\bf 5439}, 531-537.

\noindent Kannan,K., Amariglio,N., Rechavi,G., Jakob-Hirsch,J.,
Kela,I., Kaminski,N., Getz,G., Domany,E. and Givol,D. (2001) DNA
microarrays identification of primary and secondary target genes
regulated by p53, {\it Oncogene}, {\bf 20}, 2225--2234.

\noindent Kela,I. (2002) Clustering of gene expression data. M.Sc.
Thesis, Weizmann Institute (2002)

\noindent Notterman,D.A., Alon,U., Sierk,A.J. and Levine,A.J.
(2001) Transcriptional gene expression profiles of colorectal
adenoma, adenocarcinoma, and normal tissue examined by
oligonucleotide arrays, {\it Cancer Res.} {\bf 7}, 3124--3130.

\noindent Perou,C.M., Sorlie,T., Eisen,M.B., van de Rijn,M.,
Jeffrey,S.S., Rees,C.A., Pollack,J.R., Ross,D.T., Johnsen,H.,
Akslen,L.A., Fluge,O., Pergamenschikov,A., Williams,C., Zhu,S.X.,
Lonning,P.E., Borresen-Dale,A.L., Brown,P.O. and Botstein,D.
(2000) Molecular portraits of human breast tumours, {\it Nature},
{\bf 406}, 747--752.

\noindent Sorlie,T., Perou,C.M., Tibshirani,R., Aas,T.,
Geisler,S., Johnsen,H., Hastie,T., Eisen,M.B., van de Rijn,M.,
Jeffrey,S.S., Thorsen,T., Quist,H., Matese,J.C., Brown,P.O.,
Botstein,D., Eystein Lonning,P. and Borresen-Dale,A.L. (2001) Gene
expression patterns of breast carcinomas distinguish tumor
subclasses with clinical implications, {\it Proc. Natl. Acad. Sci.
USA}, {\bf 19}, 10869--10874.

\noindent Quintana,F., Getz,G., Hed,G., Domany,E., Cohen,I.R.
(2002) Cluster analysis of human autoantibody reactivities in
health and in type 1 diabetes mellitus: A bio-informatic approach
to immune complexity, {\it (unpublished)}.

}

\end{document}